\documentclass[preprint,5p,twocolumn,numbers]{elsarticle}

\usepackage{amssymb}
\usepackage{amsmath}
\usepackage[colorlinks=true, linkcolor=blue, citecolor=blue, urlcolor=blue]{hyperref}
\usepackage{balance}
\usepackage{graphicx}
\usepackage{booktabs}
\usepackage{multirow}
\usepackage{tabularx}
\usepackage{enumitem}
\usepackage{threeparttable}
\usepackage{url}
\usepackage{placeins} 

\DeclareMathOperator{\E}{E}
\DeclareMathOperator{\LL}{L}
\DeclareMathOperator{\Var}{Var}
\DeclareMathOperator{\CVE}{CVE}
\DeclareMathOperator{\CVSS}{CVSS}
\DeclareMathOperator{\Model}{M}

\newcommand{\vc}[1]{\boldsymbol{#1}}

\newcommand{\vcbeta}{\vc{\beta}}

\newcommand{\vcDelta}{\vc{\Delta}}
\newcommand{\vcone}{\textup{\textbf{\textrm{1}}}}
\newcommand{\vcx}{\textup{\textbf{\textrm{x}}}}
\newcommand{\mtx}{\textup{\textbf{\textrm{X}}}}

\binoppenalty=\maxdimen
\relpenalty=\maxdimen

\newcommand{\change}[1]{\textcolor{black}{#1}}

\begin{document}

\title{A Look at the Time Delays in CVSS Vulnerability Scoring}
%

\author[utu]{Jukka Ruohonen\corref{cor}}
\ead{juanruo@utu.fi}
\cortext[cor]{Corresponding author.}
\address[utu]{Department of Future Technologies, University of Turku, FI-20014 Turun yliopisto, Finland}

\begin{abstract}
This empirical paper examines the time delays that occur between the publication of Common Vulnerabilities and Exposures (CVEs) in the National Vulnerability Database (NVD) and the Common Vulnerability Scoring System (CVSS) information attached to published CVEs.
According to the empirical results based on regularized regression analysis of over eighty thousand archived vulnerabilities, (i) the CVSS content does not statistically influence the time delays, which, however, (ii) are strongly affected by \change{a} decreasing annual trend. In addition to these results, the paper contributes to the empirical research tradition of software vulnerabilities by a couple of insights on misuses of statistical methodology.
\end{abstract}


\maketitle

\section{Introduction}

Software vulnerabilities are software bugs that expose weaknesses in software systems. The CVSS standard is used to classify the severity of known and disclosed vulnerabilities. Once the classification and evaluation work has been completed for a vulnerability identified with a CVE, the structured and quantified severity information is stored to vulnerability databases. \change{Motivated by a recent empirical evaluation} \cite{Johnson17}, this paper examines the time delays between the publication of CVEs and the usually later publication of CVSS information. The scope is restricted to NVD and the second revision of the CVSS standard.

The use of CVSS is mandated and recommended by many state agencies for assessments in different security-critical domains \cite{Scarfone09}, including but not limited to medical devices \cite{Stine17} and the payment card industry \cite{Allodi17a}. The standard has been also incorporated into different governmental  security risk, threat, and intelligence systems. Furthermore, CVSS information is used in numerous different commercial products \cite{Johnson17}, ranging from vulnerability scanners and compliance assessment tools to automated penetration testing and intrusion detection systems. 

CVSS is also widely used in academic research. Typical application domains include risk analysis \cite{Allodi17a, Houmb10}, security audit frameworks~\cite{Aslam15}, so-called attack graphs \cite{Gallon11, MunozGonzalez17}, and empirical assessments using CVSS for different purposes~\change{\cite{Allodi17b, Morrison17, Ruohonen17TIR, Ruohonen17COMSIS}}. To these ends, a lot of work has been done to improve CVSS with different weighting algorithms~\cite{KoLee16, WangGuo12}, among other techniques \change{\cite{Geng15, Ross17}}. With some rare exceptions~\cite{Holm15}, limited attention has been given for examining how severity assessments are done in practice.

\begin{figure}[th!b]
\centering
\includegraphics[width=8cm, height=6cm]{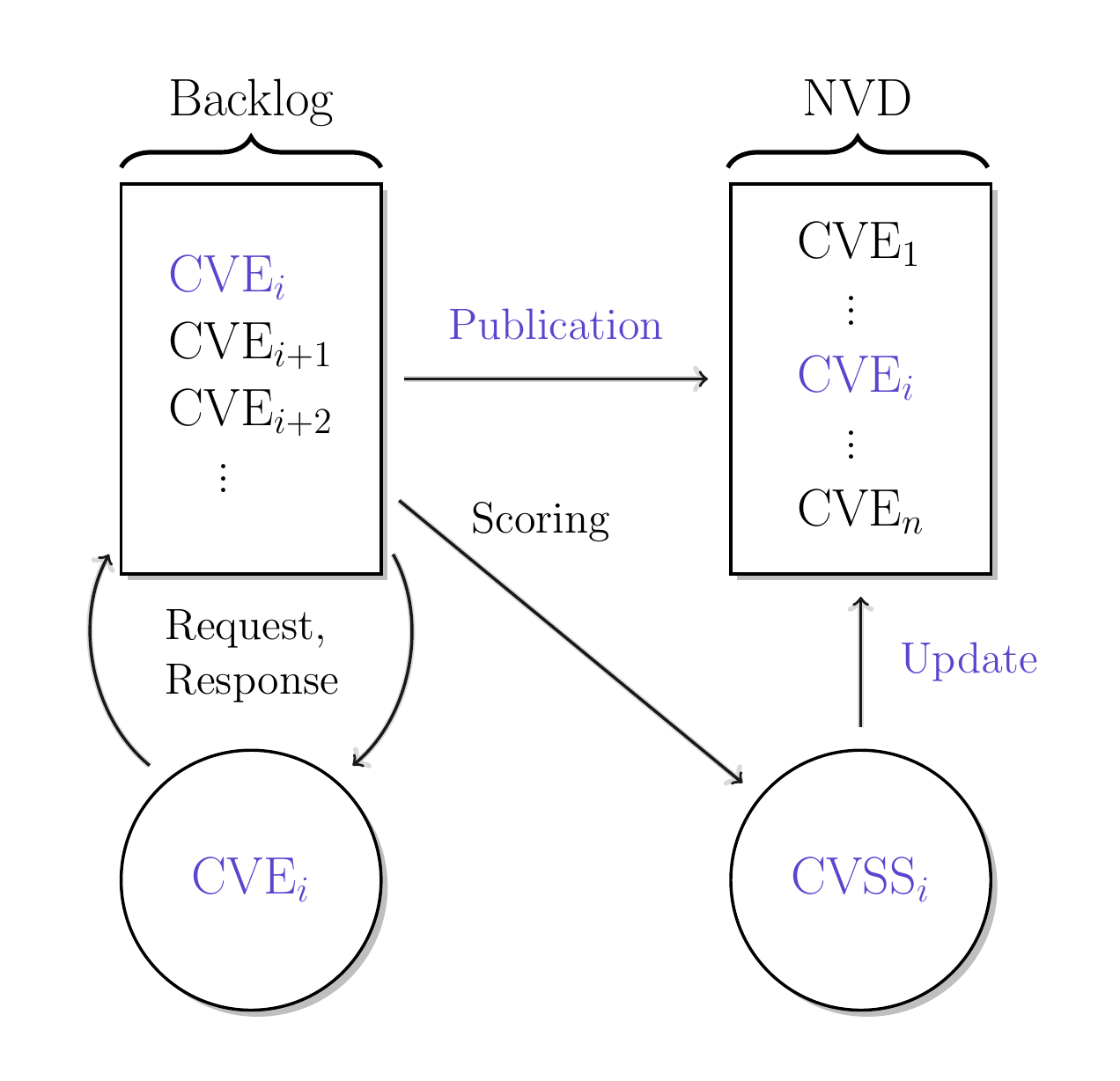}
\caption{A Simplified Model for CVSS Processing}
\label{fig: cvss processing}
\end{figure}

Practical approaches are important because CVSS has faced also challenges. Analogous to \change{problems that have affected} CVE assignments \change{\text{\cite{Ruohonen17COMSIS, Ruohonen17IWSMMENSURA}}}, different practical problems have \change{influenced} the severity assignments for CVE-stamped vulnerabilities. Excluding the actual content of the standard, the historical problems \change{related} to classification inconsistencies, time delays, and the proliferation of classification standards~\cite{EiramMartin13, Mell06}. Some of these problems have continued to exist. For instance, proliferation has continued in recent years; new standards have been introduced for classifying software misuse and configuration vulnerabilities~\cite{Alsaleh14}. \change{Some countries \cite{ZhuCao17} and companies~\cite{YonisMalaiya15} have also introduced their own} severity metrics. To examine whether also the problem with time delays is still present\change{---as has been suspected~\cite{Ladd17}}, a brief remark is required about the CVE and CVSS publication processes \change{in the context of NVD}. \change{Although the available documentation about these processes is limited \cite{NVD17b},} the sketch presented in Fig.~\ref{fig: cvss processing} is not a far-fetched \change{analytical} speculation.

The process starts when security researchers, vendors, and other related actors request CVEs for vulnerabilities they have discovered or made aware of. These request-response dynamics are handled by the non-profit MITRE corporation. As is common in software engineering, MITRE presumably maintains a backlog for the CVEs assigned, some of which may be even rejected for inclusion to NVD. Although the structure of the backlog is unknown, a simple FIFO (first-in, first-out) might be considered in order to connect the speculation to \change{a} recent theoretical work~\cite{Haldar17}. In any case, eventually the vulnerabilities accepted for archiving are published in NVD. In parallel to the coordination and archiving work related to CVEs, vulnerabilities are evaluated for their severity by the NVD team, which largely operates independently from others carrying similar evaluations~\cite{Johnson17}. Once the evaluation has been completed, the CVE-referenced vulnerability information is updated in NVD. The time lags between the initial CVE publications and the later CVSS updates constitute the empirical phenomenon examined.

There is another viewpoint to the abstract CVE backlog. \change{This viewpoint originates from the so-called switching costs, which are often high for information technology standards~\cite{Shin15}}. \change{Such theoretical costs cover also database maintenance:} even small changes made to standards may imply a lot of evaluation work particularly in case old information needs to be updated. This concern was raised also during the 2007 introduction of the second revision of the CVSS standard~\cite{Scarfone09}. In other words, updates \change{can be} costly in terms of time and resources -- given the \change{nearly ninety} thousand vulnerabilities currently archived in NVD. Therefore, it is relevant to ask the following research question (RQ) about the time lags affecting CVSS scoring.

\begin{enumerate}[label={RQ$_{\arabic{enumi}}$}]
\item{\textit{Do the time delays between CVE publications and CVSS updates vary systematically according to an annual year-to-year trend?}}\label{rq: annual}
\end{enumerate}

Another question relates to the content of the CVSS standard in terms of the vulnerabilities scored. Reflecting the disagreements among experts about the severity of some vulnerability types \cite{Holm15}, it can be hypothesized that the CVSS content itself affects the time delays. Not all vulnerabilities are equally easy (or hard) to classify in terms of severity; hence, some vulnerabilities may take a relatively short \change{(long)} time to classify. This reasoning \change{is} presented as a second research question, stated as follows.

\begin{enumerate}[label={RQ$_{\arabic{enumi}}$}, resume]
\item{\textit{Do the time delays vary systematically according to the content of the CVSS severity information?}}\label{rq: cvss}
\end{enumerate}

Finally, a third and final question can be postulated for controlling the answers to the earlier two questions:

\begin{enumerate}[label={RQ$_{\arabic{enumi}}$}, resume]
\item{\textit{Does the answer to \ref{rq: cvss} hold when also the annual trend is controlled for?}}\label{rq: cvss-annual}
\end{enumerate}

\change{According to the empirical results, only the answer to \ref{rq: annual} is positive. For predicting the time delays, the CVSS content is largely noise. The statistical effect (\ref{rq: cvss}) also fades away once the annual trend is controlled for (\ref{rq: cvss-annual}). To elaborate how these conclusions are reached}, the remainder of this paper is structured into three sections. Namely: Section~\ref{section: setup} introduces the dataset and the operationalization of the variables used, Section~\ref{section: results} outlines the statistical methodology and presents the empirical results along the way, and Section~\ref{section: discussion} finally discusses the findings.

\section{Setup}\label{section: setup}

To outline the setup for the analysis, the following discussion will address the operationalization of the delay metric examined the covariates used to model the metric.

\subsection{Response}

\change{Following the so-called vulnerability life cycle research tradition \cite{Morrison17, Ruohonen17COMSIS}}, the interest relates to a time difference
\begin{align}\label{eq: tdiff}
\Delta_i = \tau_{\CVSS_i} &- \tau_{\CVE^a_i} ,
\quad\textmd{given}
\\ \notag
\tau_{\CVSS_i} &\geq \tau_{\CVE^a_i}
\quad\textmd{for all~}i = 1, \ldots, n .
\end{align}

The integer $\tau_{\CVSS_i}$ denotes the day (timestamp) at which a CVSS entry was generated for the $i$:th CVE that was published at $\tau_{\CVE^a_i}$. In practice, the two timestamps map to the fields \texttt{cvss:generated-on-datetime} and \texttt{vuln:published-datetime} in the NVD's extensible markup language schema. Although the exact meaning of the fields is undocumented, the time differences can be interpreted as delays between CVE and CVSS publications. 

\begin{figure}[th!b]
\centering
\includegraphics[width=\linewidth, height=5cm]{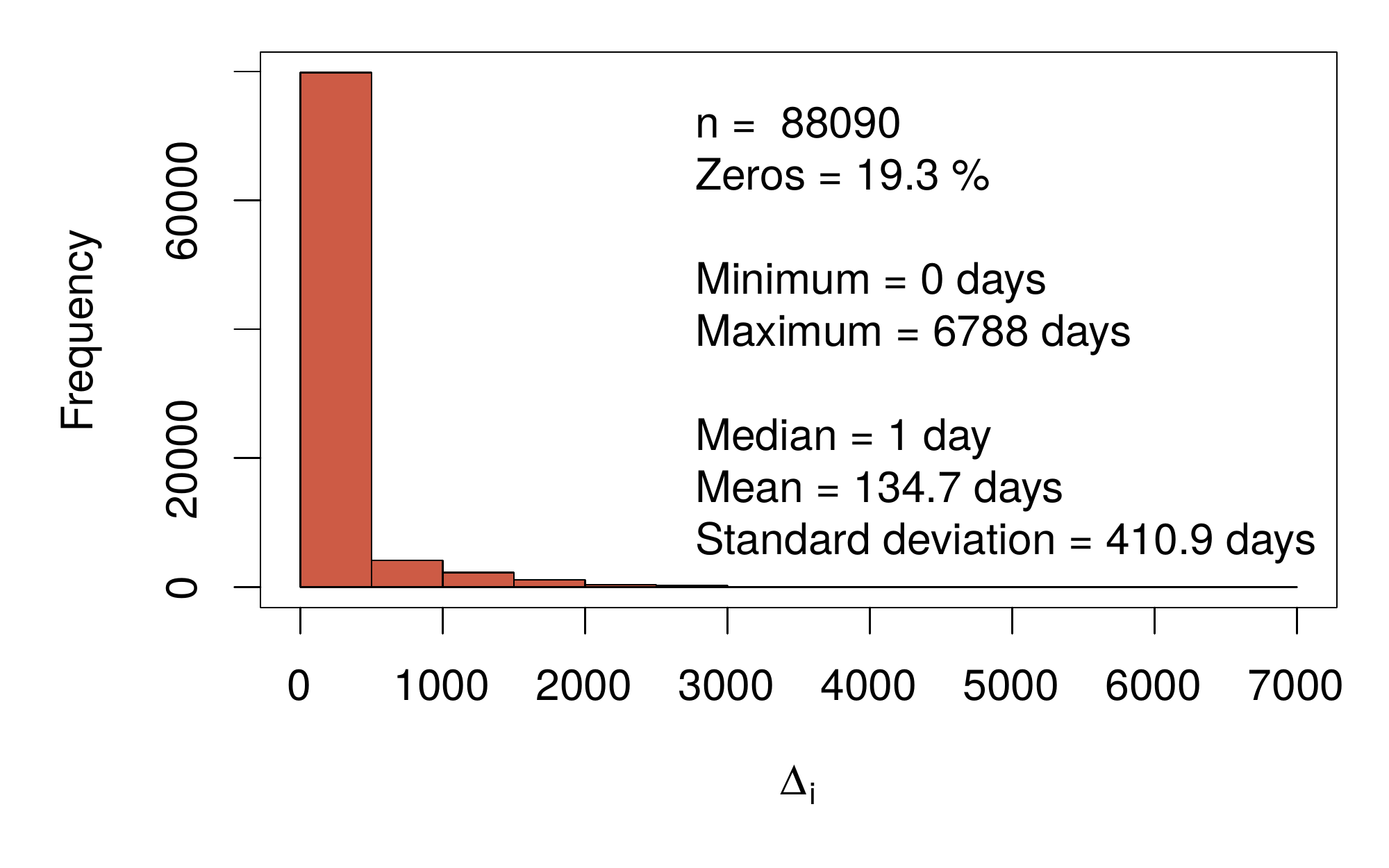}
\caption{CVE-CVSS Publication Time Delays (Eq.~\ref{eq: tdiff})}
\label{fig: dtime}
\end{figure}

Of the $89465$ archived vulnerabilities with both CVE and CVSS entries, the condition $\tau_{\CVSS_i} \geq \tau_{\CVE^a_i}$ fails to satisfy only for $1375$ vulnerabilities. Without loss of generality, these cases were excluded. The same applies to CVEs without severity records. At the time of retrieving the NVD content~\cite{NVD17a}, there were $2218$ vulnerabilities that were published but still lacked CVSS records. Most of these cases relate either to new vulnerabilities that are still in the pipeline for severity assessments, or to already published CVEs that were later rejected as inappropriate for archiving. \change{Either} way, these had to be \change{also} excluded in order for $\Delta _i$ to be defined for all cases observed. In total, the dataset examined contains $n = 89465 - 1375 = 88090$ archived cases. Given these cases, the distribution of the time delays observed is shown in Fig.~\ref{fig: dtime}.

\begin{figure}[th!b]
\centering
\includegraphics[width=\linewidth, height=5cm]{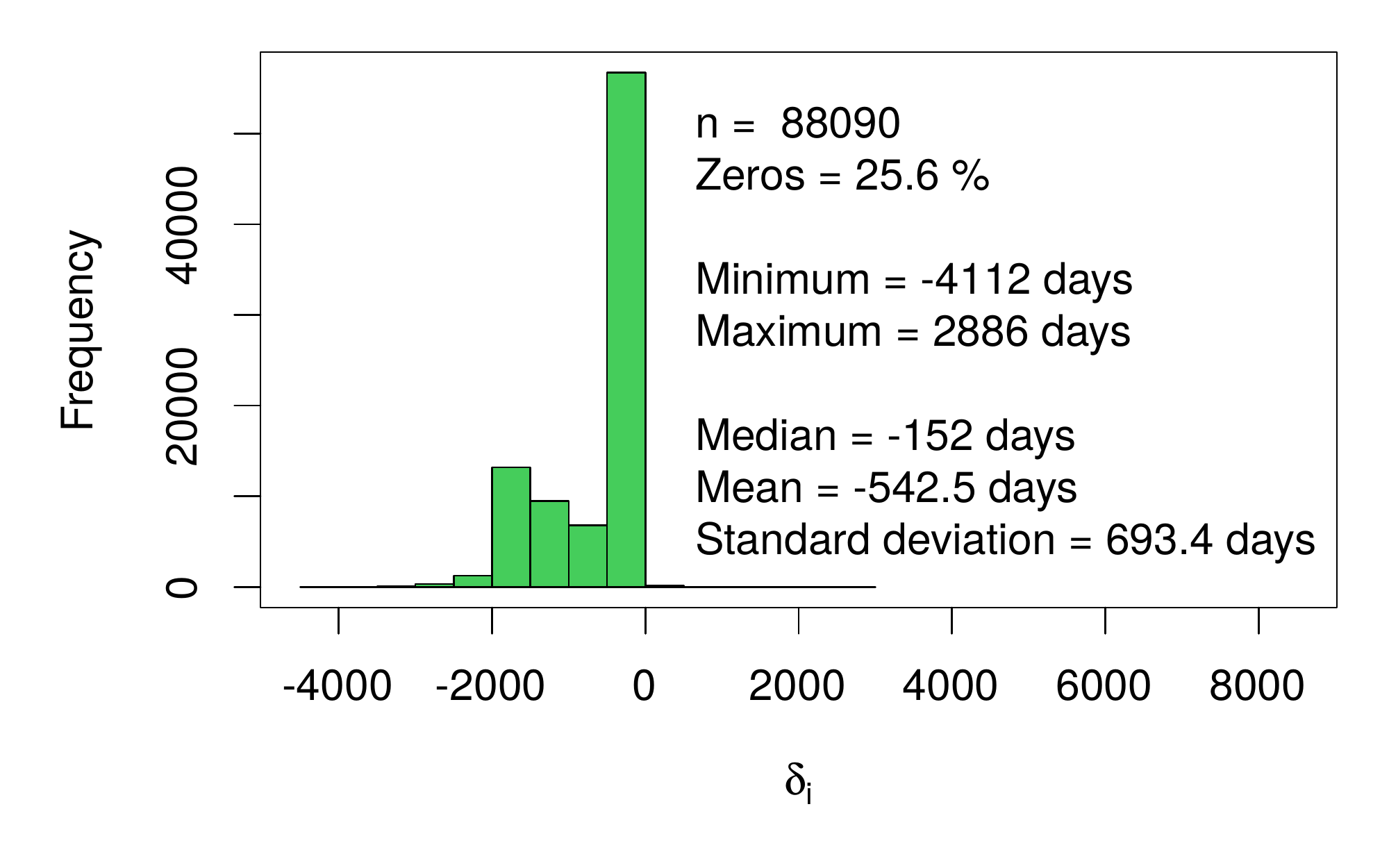}
\caption{CVE-CVSS Modification Time Delays (Eq.~\ref{eq: mtime})}
\label{fig: mtime}
\end{figure}

The timelines exhibit a heavy-tailed distribution with extremely long right tail. A half of the vulnerabilities observed have seen severity assignments already a day after CVEs were published, but the standard deviation is still over a year. Most of this deviation is caused by a few extreme outliers for which the severity scores were assigned even a decade after the CVEs were originally published. 

To briefly probe these outliers further, Fig.~\ref{fig: mtime}. displays the distribution of another time difference
\begin{align}\label{eq: mtime}
\delta_i = \tau_{\CVSS_i} &- \tau_{\CVE^b_i} ,
\end{align}
where $\tau_{\CVE^b_i}$ denotes the \texttt{vuln:last-modified-datetime} field in NVD. The large amount of negative values indicate that CVEs are often updated after these were already published with CVSS information. Interestingly, $187$ outlying cases satisfy $\delta_i > 0$, which may point toward some inconsistencies in database maintenance; CVSS information was generated without updating the corresponding $\tau_{\CVE^b_i}$ timestamps. About a quarter of the cases observed satisfy $\delta_i = 0$, meaning that the latest CVE modifications matched the generation of severity information.

\subsection{Covariates}

Two types of covariates are used for modeling \change{the time delays in \eqref{eq: tdiff}}. The first contains the CVSS information itself. The CVSS (v.~2) standard \cite{FIRST07} classifies the impact of vulnerabilities according to confidentiality, integrity, and availability (CIA). Each letter in the CIA acronym further expands into three categories that characterize the impact upon successfully exploiting the vulnerability in question. \change{Thus, the} analytical structure behind the impact dimension can be illustrated with a diagram:

\begin{scriptsize}
\begin{align*}
\textmd{IMPACT} \in
\begin{cases}
\textmd{CONFIDENTIALITY} 
&\in \begin{cases}
\textmd{NONE$^*$} \\
\textmd{PARTIAL} \\
\textmd{COMPLETE} \\
\end{cases}
\\
\textmd{INTEGRITY}
&\in \begin{cases}
\textmd{NONE$^*$} \\
\textmd{PARTIAL} \\
\textmd{COMPLETE} \\
\end{cases}
\\
\textmd{AVAILABILITY}
&\in \begin{cases}
\textmd{NONE$^*$} \\
\textmd{PARTIAL} \\
\textmd{COMPLETE} \\
\end{cases}
\end{cases}
\end{align*}
\end{scriptsize}

The three impact metrics measure the severity of a vulnerability on a system after the vulnerability has already been exploited. However, not all vulnerabilities can be exploited; therefore, the CVSS standard specifies also an exploitability dimension for vulnerabilities. Like with the impact dimension, exploitability expands into three metrics (access vector, complexity, and authentication) that can each take three distinct values. The analytical meaning can be again summarized with the following diagram:

\begin{scriptsize}
\begin{align*}
\textmd{EXPLOITABILITY} \in
\begin{cases}
\textmd{VECTOR} 
&\in \begin{cases}
\textmd{LOCAL$^*$} \\
\textmd{NETWORK} \\
\textmd{ADJACENT} \\
\end{cases}
\\
\textmd{COMPLEXITY}
&\in \begin{cases}
\textmd{LOW$^*$} \\
\textmd{MEDIUM} \\
\textmd{HIGH} \\
\end{cases}
\\
\textmd{AUTHENTICATION}
&\in \begin{cases}
\textmd{NONE$^*$} \\
\textmd{SINGLE} \\
\textmd{MULTIPLE} \\
\end{cases}
\end{cases}
\end{align*}
\end{scriptsize}

The rationale for the impact and exploitability metrics relate to different combinatory relationships between the different values the metrics can take. For instance, it is \change{probable} that mass-scale attacking tools target \change{less complex} vulnerabilities that can be exploited through a network without performing authentication, possibly regardless of the impact upon confidentiality, integrity, and availability. There \change{exists} also some empirical evidence along these lines \cite{Allodi17b}. However, the impact and exploitability dimensions both relate to intrinsic characteristics of vulnerabilities; they are constant across time and environments. For instance, EXPLOITABILITY cannot answer to a temporal question about whether an exploit is known to exists for the vulnerability in question \change{\cite{Ross17, YonisMalaiya15}}. The same point extends toward NVD in general \cite{Garcia14}. For these and other reasons, the new (v.~3) standard for CVSS enlarges the dimensions toward temporal and environmental metrics. 

For the present purposes, however, the impact and exploitability dimensions are sufficient for soliciting an answers to \ref{rq: cvss}. \change{This choice is also necessitated by the paper's focus on NVD, which does not currently provide full CVSS~v.~3 information \cite{NVD17c}. Despite of this limitation, a correlation} between the six CVSS metrics and $\Delta_i$ could be expected due to the fairly detailed criteria used for the manual classification. Complex \change{vulnerabilities with severe impact} may require more evaluation work than trivial vulnerabilities; a remote buffer overflow vulnerability is usually more difficult to interpret compared to a trivial cross-site scripting vulnerability. \change{Also the reverse direction is theoretically possible; more effort may be devoted for high-profile vulnerabilities~\cite{Ladd17}. Either way, \ref{rq: cvss} seems like a sensible hypothesis worth asking.}

With regard to statistical modeling, the three impact metrics and the three exploitability metrics are included in the models as so-called dummy variables. For each metric, the reference category is marked with a star in the previous two diagrams. For instance, INTEGRITY is expanded into two dummy variables, INTEGRITY(PARTIAL) and INTEGRITY(COMPLETE), say, the effects of which are compared against INTEGRITY(NONE), which cannot be included in the models due to multicollinearity. The same strategy applies to the metrics used for evaluating \ref{rq: annual}. Namely, the annual effects are proxied through $18$ dummy variables that record the year at which a vulnerability was published according to $\tau_{\CVE^a_i}$. Because only five vulnerabilities were published in the 1980s and a negiligle amount (about 1.8~\%) in the 1990s, the reference category for the annual dummy variables is formed by collapsing all vulnerabilities published prior to 2000 into a single group. Given the two CVSS dimensions and the dummy variable approximation for the annual trend, three model matrices ($\mtx_1$, $\mtx_2$, and $\mtx_3$) are used in the statistical computation:
\begin{equation}\label{eq: model matrices}
\begin{cases}
\Model_1:~\mtx_1 = 
\bigl[ \vcone, \mtx_\textmd{IMPACT} \bigl],  \\
\Model_2:~\mtx_2 = 
\bigl[ \mtx_1, \mtx_\textmd{EXPLOITABILITY}  \bigl], \\
\Model_3:~\mtx_3 = 
\bigl[ \mtx_2, \mtx_\textmd{ANNUAL}  \bigl] .
\end{cases}
\end{equation}

The first model $\Model_1$ regresses $\vcDelta = [ \Delta_1, \ldots, \Delta_n ]^\prime$ against a constant represented by a $n$-length vector of ones, $\vcone$, and the six impact dummy variables present in the $(n \times 6)$ matrix $\mtx_\textmd{IMPACT}$. The second model is identical except that further six dummy variables are included for measuring the exploitability dimension. The third and final model includes all information used. 

\begin{figure}[th!b]
\centering
\includegraphics[width=\linewidth, height=7cm]{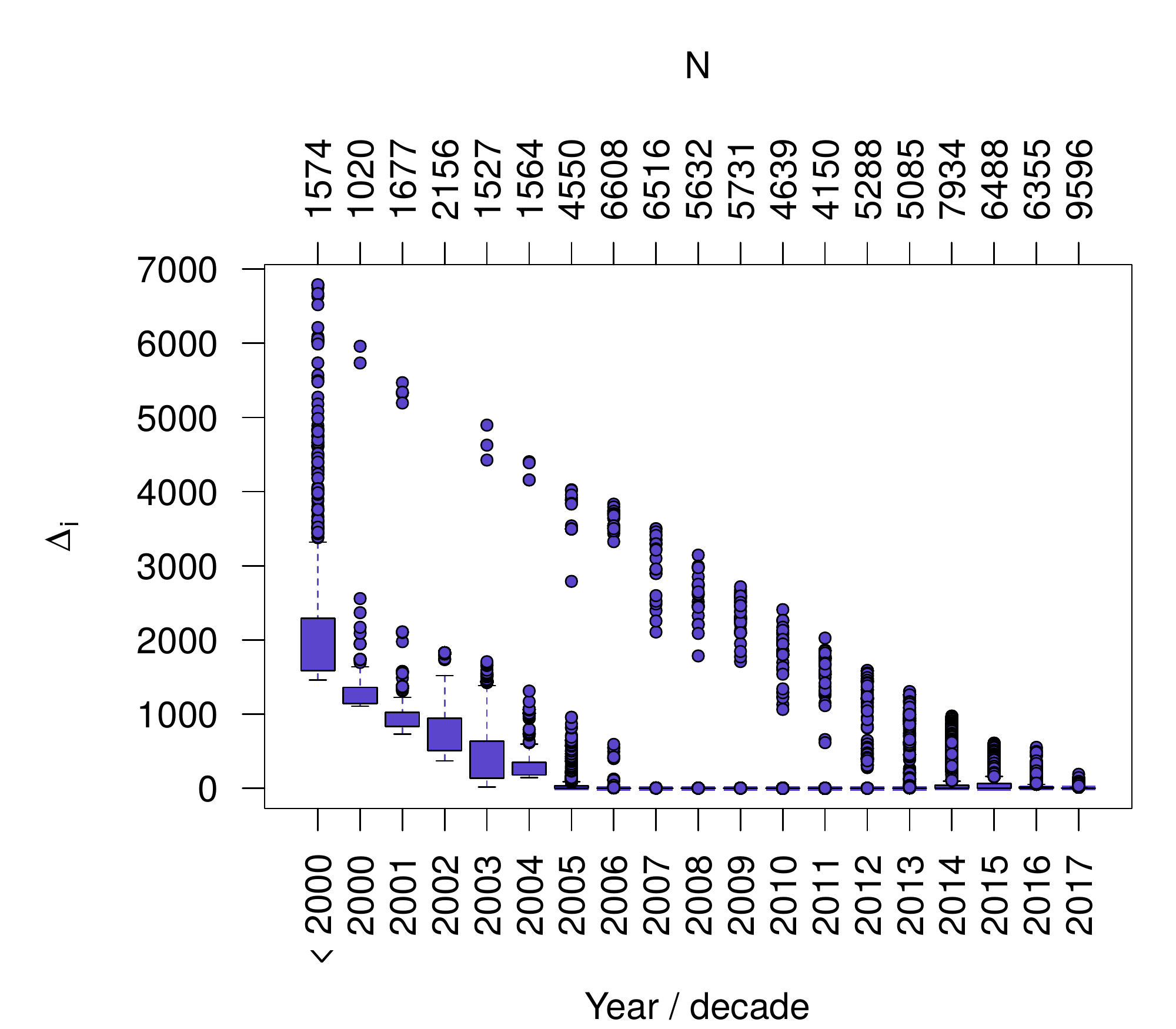}
\caption{Annual Time Delays (based on $\tau_{\CVE^a_i}$)}
\label{fig: annual}
\end{figure}

Despite of the growing number of CVEs processed from the circa mid-2000s onward \cite{Ruohonen16GIQ}, the time delays for CVSS processing have steadily decreased over the years. As can be seen from Fig.~\ref{fig: annual}, there have been no extreme outliers in recent years, meaning that most of the right tail in Fig.~\ref{fig: dtime} is attributable to older CVEs. A possible but speculative explanation is that the work done to update old CVEs with CVSS (v.~2) information has mostly been completed. 

The strong decreasing trend is likely to support a positive answer to the research question \ref{rq: annual}. Given this prior expectation, the main interest in the forthcoming analysis relates to the statistical effect of the impact and exploitability metrics when also the annual trend is modeled. One strategy for evaluating the research question \ref{rq: cvss-annual} is to compare the models $\Model_1$ and $\Model_2$ against the full information model $\Model_3$. If the CVSS metrics provide statistical power for predicting $\vcDelta$, this power should be visible also when the decreasing annual trend is controlled for.

\section{Results}\label{section: results}

The response $\vcDelta$ represents a count data vector; each observation in the vector counts the days between CVE and CVSS publications in NVD. Thus, a Poisson regression model provides a natural starting point for modeling the time delays. The expected value of the response thus is
\begin{equation}\label{eq: conditional mean poisson}
\E\left(\vcDelta~\vert~\mtx_j\right)
= e^{\mtx_j \vcbeta} ,
\end{equation}
where $\mtx_j$ is a given model matrix from~\eqref{eq: model matrices} and $\vcbeta$ a \text{$k$-length} vector of regression coefficients, including the intercept $\beta_1$. This conditional mean is always positive. 

However, the model assumes that $\vcDelta$ is distributed from the Poisson distribution, which, in turn, implies that the mean of the time delays should equal the variance of the delays. As can be concluded from the numbers shown in Fig.~\ref{fig: dtime}, this assumption is clearly problematic in the current setting. While $\vcbeta$ is still consistently estimated, the apparent overdispersion, $\Var(\vcDelta) > \E(\vcDelta)$, affects the standard errors of the regression coefficients, and, hence, the statistical significance of the coefficients. A common solution to tackle the overdispersion problem is to estimate a so-called negative binomial model (NBM) instead, although the conventional ordinary least squares (OLS) regression often works well in applied problems when the response is suitably transformed~\cite{Ives15}. Thus, instead of \eqref{eq: conditional mean poisson}, consider that the conditional mean is given by an OLS regression
\begin{equation}\label{eq: conditional mean ols}
\E(\ln[\vcDelta + 1]~\vert~\mtx_j)
= \E(\tilde{\vcDelta}~\vert~\mtx_j)
= \mtx_j \vcbeta ,
\end{equation}
such that
\begin{equation}\label{eq: rss ols}
\hat{\vcbeta}_a =
\min_{\vcbeta} 
(\tilde{\vcDelta} - \mtx_j\vcbeta)^\prime
(\tilde{\vcDelta} - \mtx_j\vcbeta)  .
\end{equation}

When applied to the full model matrix $\mtx_3$, the adjusted coefficient of determination is 0.64 for this OLS regression. In other words, the general model performance is quite decent, given the limited amount of information used to model the severity assignment timelines. Moreover, only three coefficients in $\hat{\vcbeta}_a$ are not significant at the conventional $p < 0.05$ threshold. By further testing the joint significance of the dummy variable groups with a \text{$F$-test}, all groups are significant at a $p < 0.001$ level. Also the combined forward-stepwise and backward-stepwise algorithm (as implemented in the \texttt{step} function for R) retains all coefficients in $\hat{\vcbeta}_a$. As is common in applied problems \cite{Rydberg17}, the $\tilde{\vcDelta} = \ln(\vcDelta + 1)$ transformation does not account for the high positive skew; therefore, another test can be computed by using an R implementation \cite{Zeileis04} for a consistent covariance matrix estimator \cite{White80}. However, the results do not diverge much from the plain OLS estimates; only one additional coefficient is insignificant at a $p < 0.05$ threshold. Finally, analogous conclusions can be reached by estimating a negative binomial regression model with the assumption that \change{$\Var(\vcDelta) = \E(\vcDelta) + \phi[\E(\vcDelta)]^2$}, where $\phi$ is a parameter to be estimated \change{\cite{Lawless87, WeiLovegrove13}}. By again using an R implementation~\cite{aod}, only two coefficients attain $p \geq 0.05$.

\begin{figure}[th!b]
\centering
\includegraphics[width=\linewidth, height=5cm]{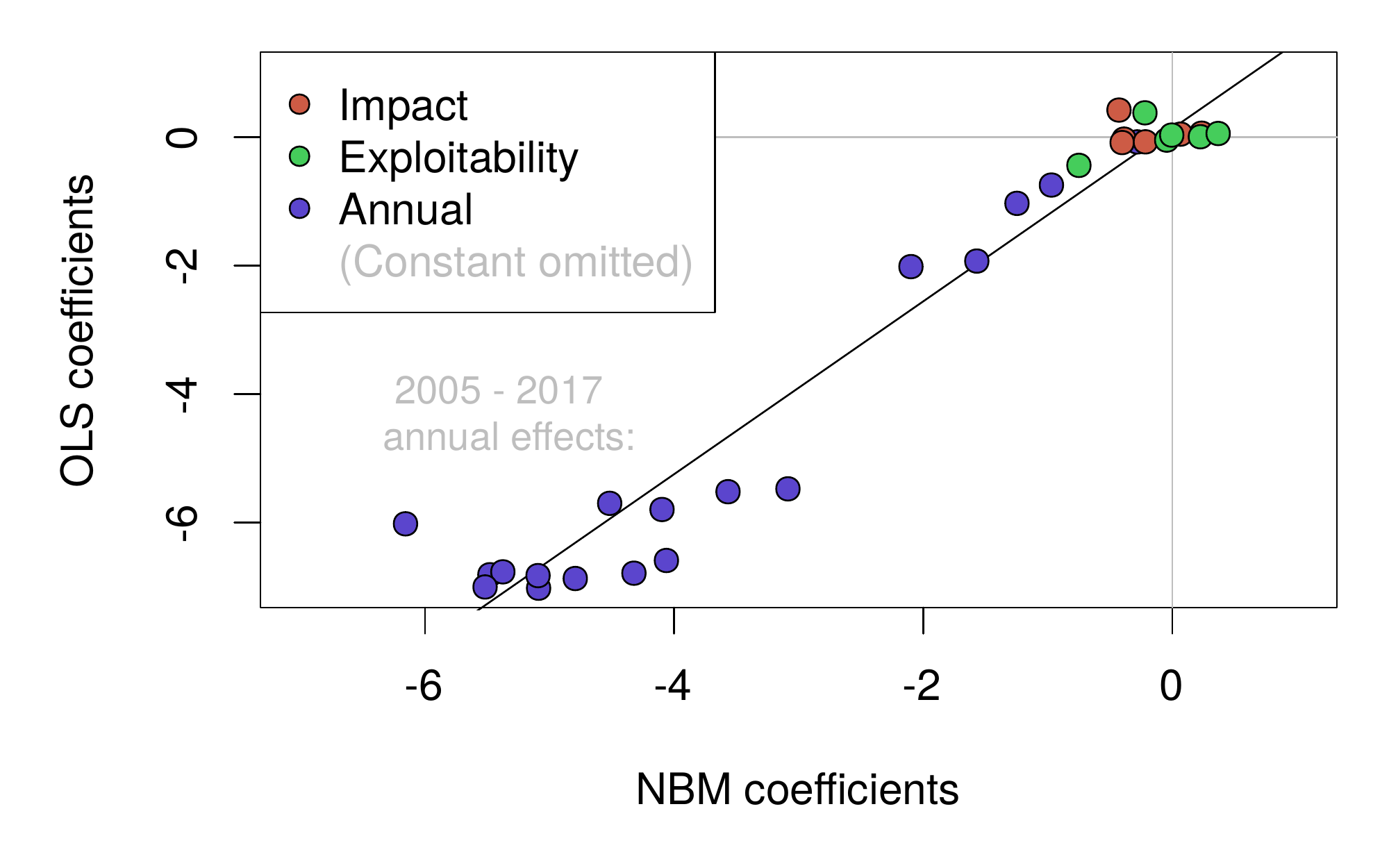}
\caption{Coefficients from the OLS and NBM Regressions ($\Model_3$)}
\label{fig: coef ols negbin}
\end{figure}

Thus, based on statistical significance, positive answers would be given to all three research questions. This conclusion would be unwarranted, however. Most of the coefficients in the $\Model_3$ model are close to zero, irrespective of the estimation strategy. Since all covariates are dummy variables (and, hence, have the same scale), this observation can be illustrated in the form of Fig.~\ref{fig: coef ols negbin}, which plots the OLS coefficients \text{($y$-axis)} against the corresponding NBM coefficients \text{($x$-axis)}, omitting the constant $\hat{\beta}_1$. As can be seen, there are some differences between the two regression coefficient vectors, but these differences apply mostly to the annual effects. In particular, the coefficients for the impact and exploitability dimensions are very close to zero without notable differences between the OLS and NBM estimates. The largest absolute coefficient values are obtained for the annual effects from 2005 to 2017. These coefficients exhibit also the largest differences between the OLS and the negative binomial estimates. 

To examine these observations further, the so-called least absolute shrinkage and selection operator (LASSO) provides a good tool. The LASSO method is a regression model that uses regularization in order to improve prediction accuracy and  feature selection. When compared to other regularized regression models, such as the so-called Ridge regression, LASSO can shrink some coefficients exactly to zero. Although the feature selection properties are not entirely ideal for hypothesis testing \cite{LiSillanpaa12}, this property is desirable for further examining whether regularization pushes the coefficients for all of the CVSS metrics toward zero. It should be noted that dropping individual dummy variables based on feature selection is usually unwarranted because interpretation of the coefficients changes---but if all of the impact and exploitability dummy variables are regularized toward zero, there is not much to interpret. If this is the case, there is also no particular reason to consider more complex estimation strategies, such as the so-called group LASSO method~\cite{Vidaurre13}. A brief elaboration is required also about the more classical LASSO regressions.

Instead of minimizing the residual sum of squares in \eqref{eq: rss ols}, LASSO minimizes penalized sum of squares given by
\begin{equation}\label{eq: lasso ols}
\hat{\vcbeta}_b =
\min_{\vcbeta}
\left\lbrace
\frac{1}{2n}
\sum^n_{i=1} 
(\tilde{\Delta}_i - \vcx_{ji}^\prime\vcbeta)^2
+ \lambda \sum^k_{s=2} \vert \beta_s \vert
\right\rbrace ,
\end{equation}
where $\lambda \geq 0$ is known as the shrinkage factor, and the scaling by $(1/2n)$ is done to ease comparisons with different sample sizes \cite{Hastie15}. The penalty is given by the $L_1$ norm, that is, the sum of the absolute coefficient values, omitting the constant present in $\mtx_j$. If $\lambda$ is zero, the solution reduces to the OLS estimates, and when $\lambda \to \infty$, all coefficients in $\hat{\vcbeta}_b$ tend to zero. Despite of the overdispersion, the Gaussian LASSO in~\eqref{eq: lasso ols} can be accompanied with a Poisson LASSO as an additional robustness check.

\begin{figure}[p!]
\centering
\includegraphics[width=\linewidth, height=22cm]{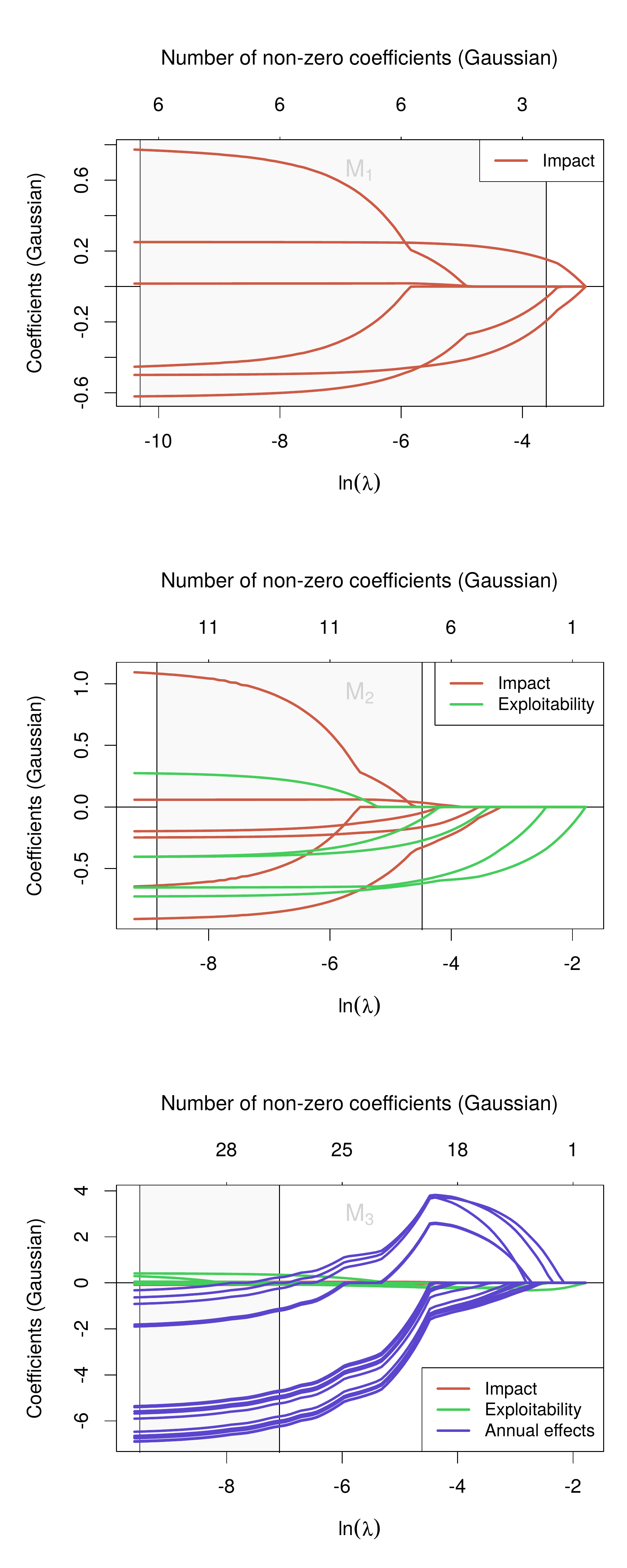}
\caption{\change{Gaussian LASSO Estimates} ($\hat{\vcbeta}_b$)}
\label{fig: lasso ols}
\end{figure}

\begin{figure}[p!]
\centering
\includegraphics[width=\linewidth, height=22cm]{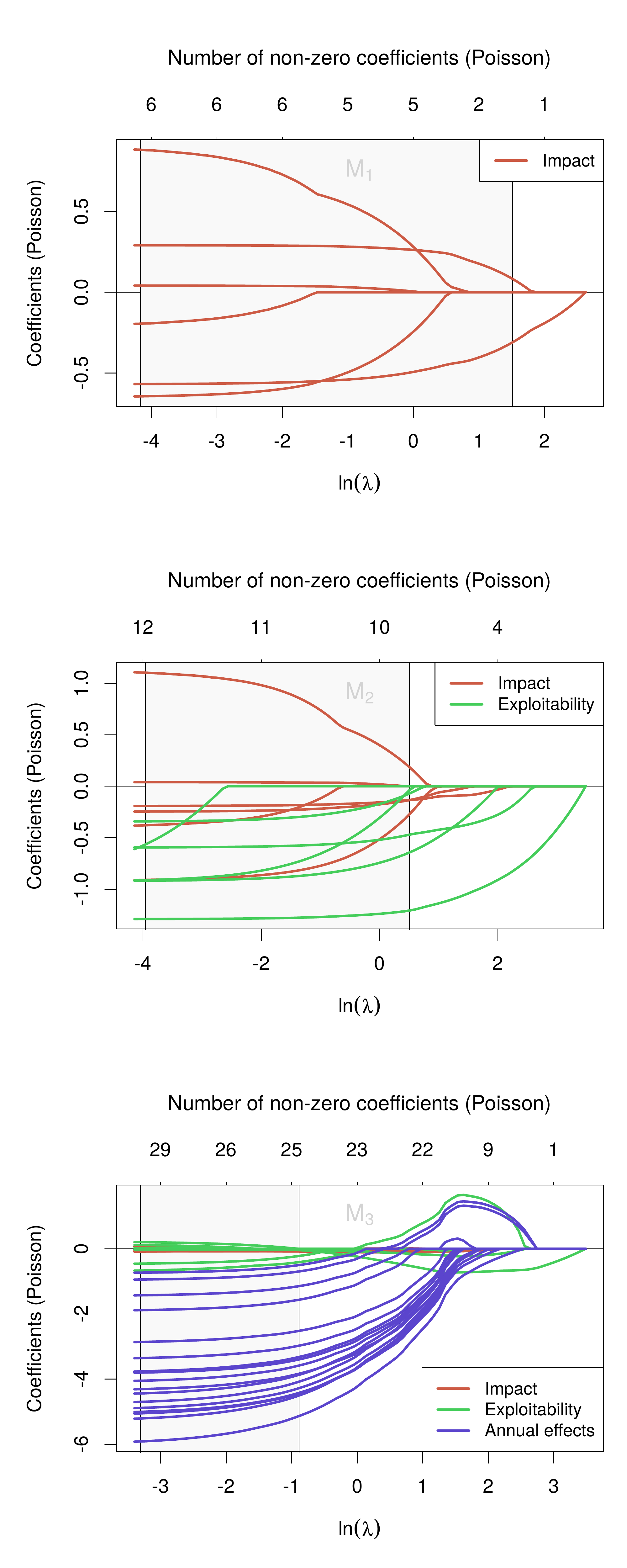}
\caption{\change{Poisson LASSO Estimates} ($\hat{\vcbeta}_c$)}
\label{fig: lasso poisson}
\end{figure}

The so-called quasi log-likelihood for Poisson regression \change{can be} obtained by left-multiplying the logarithm of the expected values in \eqref{eq: conditional mean poisson} by $\vcDelta$ and subtracting $\E\left(\vcDelta~\vert~\mtx_j\right)$ from the result \cite{McCullagh83}. Given this quasi log-likelihood,
\begin{equation}
\LL\left(\vcbeta~\vert~\vcDelta, \mtx_j\right) 
= \vcDelta\mtx_j\vcbeta - \exp(\mtx_j\vcbeta) ,
\end{equation}
LASSO optimizes 
\begin{equation}
\hat{\vcbeta}_c = 
\min_{\vcbeta} \left\lbrace
- \frac{\LL\left(\vcbeta~\vert~\vcDelta, \mtx_j\right)}{n} + \lambda \sum^k_{s=2} \vert \beta_s \vert 
\right\rbrace ,
\end{equation}
for the Poisson regression \cite{Hastie15}. By again using an R implementation \cite{glmnet14}, the results from the LASSO computations are shown in Figures~\ref{fig: lasso ols} and \ref{fig: lasso poisson} for the Gaussian and Poisson specifications. The coefficient magnitudes are shown in the $y$-axes, the lower $x$-axes represent different values of $\lambda$ in logarithm scale, and the upper $x$-axes denote the number of coefficients not regularized to zero. The shaded region is based on a $10$-fold cross-validation: in each plot, the left endpoint of the region corresponds with the value of $\lambda$ that gives the minimum cross-validation error, while the right endpoint is one standard error from this minimum.

In both figures, the models $\Model_1$ and $\Model_2$ yield large absolute coefficient magnitudes for the CVSS metrics. Furthermore, the coefficients retain their magnitudes rather long as the shrinkage factor increases. For instance, the upper-left plot indicates that none of the impact metrics are regularized to zero in the Gaussian specification until about $\lambda = \exp(-6)$. However, when the annual affects are included in $\Model_3$, all of the CVSS metrics are very close to zero particularly with respect to $\hat{\vcbeta}_b$. Although a couple of exploitability metrics retain their magnitudes within the cross-validation region shown in the lower-right plot in Fig.~\ref{fig: lasso poisson}, the same conclusion applies more or less also to the Poisson LASSO model. Furthermore, within the cross-validation regions, both $\hat{\vcbeta}_b$ and $\hat{\vcbeta}_c$ compare well to the OLS and NBM coefficient vectors illustrated in Fig.~\ref{fig: coef ols negbin}. To conclude: when predicting the time delay from CVE publications to CVSS assignments, the actual CVSS content is \change{largely} noise; the most relevant readily available information comes with the decreasing annual trend.

\section{Discussion}\label{section: discussion}

This short empirical paper examined the time delays that affect CVSS scoring work in the context of NVD. Three research questions were presented for guiding the empirical analysis based on regression methods. The results are easy to summarize. The CVSS content is correlated with the time delays (\ref{rq: cvss}), but the correlations are spurious; the decreasing annual trend affecting the time delays (\ref{rq: annual}) also makes the effects of the CVSS content negiligle (\ref{rq: cvss-annual}). Three points are worthwhile to raise about the significance of these empirical findings.

First, the negative answers to \ref{rq: cvss} and \ref{rq: cvss-annual} are positive findings in terms of practical applications using CVSS information. Whether the application context is governmental security intelligence systems or commercial security assessment tools, there is currently no particular reason to worry that a NVD data feed would show significant delays for the CVSS information. Likewise, in 2017, there is no reason to suspect that information for severe vulnerabilities would tend to arrive \change{later (or earlier)} than information for mundane vulnerabilities. However, this conclusion does not apply to historical contexts, and, moreover, the historically long delays affect also academic research.

Second, the positive answer to \ref{rq: annual} is a negative finding in terms of existing academic research; the historically long time delays presumably translate into selection biases in some existing empirical studies using CVSS information. Without naming any particular academic study, consider that a hypothetical article published in the late 2000s used a NVD-based dataset of CVE-referenced vulnerabilities published between 2000 and 2007, say. The long time delays during this period imply that a lot of the vulnerabilities in the dataset could not have had CVSS information. Consequently, some existing academic studies are exposed to difficult questions related to sample selection and missing values, among other issues. This concern is particularly pronounced regarding studies that examine time-sensitive topics such as vulnerability disclosure.

Third, the results echo the recently raised concern about the misuse of statistical significance in the \change{software vulnerability context} \cite{Massacci17}. It seems that the size of archival material stored to vulnerability databases has surpassed a point after which statistical significance starts to lose its usefulness for inference in applied research. The current rate of new vulnerabilities archived -- about 17 per day in 2016 -- implies that the problem with statistical significance is only going to get worse. The point is particularly important in case CVEs are referenced \change{with} other datasets, including big data \change{outputted by} intrusion detection \change{and related} systems. The regularized regression models used in this paper offer one solution to consider \change{in further applications, but more} research is required to assess the existing biases and the potential means for moving forward.

\balance
\bibliographystyle{apalike}

\begin{thebibliography}{}

\bibitem[Allodi and Massacci, 2017a]{Allodi17b}
Allodi, L. and Massacci, F. (2017a).
\newblock {A}ttack {P}otential in {I}mpact and {C}omplexity.
\newblock In {\em Proceedings of the International Conference on Availability,
  Reliability and Security (ARES 2017)}, pages 32:1--32:6, Reggio Calabria.
  ACM.

\bibitem[Allodi and Massacci, 2017b]{Allodi17a}
Allodi, L. and Massacci, F. (2017b).
\newblock {S}ecurity {E}vents and {V}ulnerability {D}ata for {C}ybersecurity
  {R}isk {E}stimation.
\newblock {\em Risk Analysis}, 37(8):1606--1627.

\bibitem[Alsaleh and {Al-Shaer}, 2014]{Alsaleh14}
Alsaleh, M.~N. and {Al-Shaer}, E. (2014).
\newblock {E}nterprise {R}isk {A}ssessment {B}ased on {C}ompliance {R}eports
  and {V}ulnerability {S}coring {S}ystems.
\newblock In {\em Proceedings of the Workshop on Cyber Security Analytics,
  Intelligence and Automation (SafeConfig 2014)}, pages 25--28, Scottsdale.
  ACM.

\bibitem[Aslam et~al., 2015]{Aslam15}
Aslam, M., Gehrmann, C., and Bj\"orkman, M. (2015).
\newblock {ASArP}: {A}utomated {S}ecurity {A}ssessment \& {A}udit of {R}emote
  {P}latforms {U}sing {TCG-SCAP} {S}ynergies.
\newblock {\em Journal of Information Security and Applications}, 22:28--39.

\bibitem[Eiram and Martin, 2013]{EiramMartin13}
Eiram, C. and Martin, B. (2013).
\newblock {T}he {CVSSv2} {S}hortcomings, {F}aults, and {F}ailures
  {F}ormulation.
\newblock {R}isk {B}ased {S}ecurity and the {O}pen {S}ecurity {F}oundation
  {(OSF)}. Available online in September 2017:
  \url{http://www.riskbasedsecurity.com/reports/CVSS-ShortcomingsFaultsandFailures.pdf}.

\bibitem[FIRST, 2007]{FIRST07}
FIRST (2007).
\newblock {A} {C}omplete {G}uide to the {C}ommon {V}ulnerability {S}coring
  {S}ystem {V}ersion {2.0}, {FIRST.ORG}.
\newblock Available online in June 2015:
  \url{https://www.first.org/cvss/cvss-v2-guide.pdf}.

\bibitem[Gallon and Bascou, 2011]{Gallon11}
Gallon, L. and Bascou, J.-J. (2011).
\newblock {CVSS} {A}ttack {G}raphs.
\newblock In {\em Proceedings of the Seventh International Conference on Signal
  Image Technology \& Internet-Based Systems (SITIS 2011)}, pages 24--31,
  Dijon. IEEE.

\bibitem[Garcia et~al., 2014]{Garcia14}
Garcia, M., Bessani, A., Gashi, I., Neves, N., and Obelheiro, R. (2014).
\newblock {A}nalysis of {O}perating {S}ystem {D}iversity for {I}ntrusion
  {T}olerance.
\newblock {\em Software: Practice and Experience}, 44(6):735--770.

\bibitem[Geng et~al., 2015]{Geng15}
Geng, J., Ye, D., and Luo, P. (2015).
\newblock {P}redicting {S}everity of {S}oftware {V}ulnerability {B}ased on
  {G}rey {S}ystem {T}heory.
\newblock In {\em Proceedings of the International Conference on Algorithms and
  Architectures for Parallel Processing (ICA3PP), Lecture Notes in Computer
  Science (Volume 9532)}, pages 143--152, Zhangjiajie. Springer.

\bibitem[Haldar and Mishra, 2017]{Haldar17}
Haldar, K. and Mishra, B.~K. (2017).
\newblock {M}athematical {M}odel on {V}ulnerability {C}haracterization and
  {I}ts {I}mpact on {N}etwork {E}pidemics.
\newblock {\em International Journal of System Assurance Engineering and
  Management}, 8(2):378--392.

\bibitem[Hastie and Qian, 2014]{glmnet14}
Hastie, T. and Qian, J. (2014).
\newblock {G}lmnet {V}ignette.
\newblock {A}vailable online in September 2017:
  \url{https://web.stanford.edu/~hastie/glmnet/glmnet_alpha.html}.

\bibitem[Hastie et~al., 2015]{Hastie15}
Hastie, T., Tibshirani, R., and Wainwright, M. (2015).
\newblock {\em {S}tatistical {L}earning with {S}parsity: {T}he {L}asso and
  {G}eneralizations}.
\newblock CRC Press, Taylor \& Francis, Boca Raton.

\bibitem[Holm and Afridi, 2015]{Holm15}
Holm, H. and Afridi, K.~K. (2015).
\newblock {A}n {E}xpert-{B}ased {I}nvestigation of the {C}ommon {V}ulnerability
  {S}coring {S}ystem.
\newblock {\em Computers \& Security}, 53:\text{18--30}.

\bibitem[Houmb et~al., 2010]{Houmb10}
Houmb, S.~H., Franqueira, V. N.~L., and Engum, E.~A. (2010).
\newblock {Q}uantifying {S}ecurity {R}isk {L}evel from {CVSS} {E}stimates of
  {F}requency and {I}mpact.
\newblock {\em Journal of Systems and Software}, 83:\text{1622--1634}.

\bibitem[Ives, 2015]{Ives15}
Ives, A.~R. (2015).
\newblock {F}or {T}esting the {S}ignificance of {R}egression {C}oefficients,
  {G}o {A}head and {L}og-{T}ransform {C}ount {D}ata.
\newblock {\em Methods in Ecology and Evolution}, 6(7):828--835.

\bibitem[Johnson et~al., 2017]{Johnson17}
Johnson, P., Lagerstr\"om, R., Ekstedt, M., and Franke, U. (2017).
\newblock {C}an the {C}ommon {V}ulnerability {S}coring {S}ystem be {T}rusted?
  {A} {B}ayesian {A}nalysis.
\newblock {\em IEEE Transactions on Dependable and Secure Computing}.
\newblock Published online in December 2016.

\bibitem[Ko et~al., 2016]{KoLee16}
Ko, J., Lee, S., and Shon, T. (2016).
\newblock {T}owards a {N}ovel {Q}uantification {A}pproach {B}ased on {S}mart
  {G}rid {N}etwork {V}ulnerability {S}core.
\newblock {\em International Journal of Energy Research}, 40(3):298--312.

\bibitem[Ladd, 2017]{Ladd17}
Ladd, B. (2017).
\newblock {T}he {R}ace {B}etween {S}ecurity {P}rofessionals and {A}dversaries.
\newblock {R}ecorded {F}uture {B}log. {A}vailable online in November 2017:
  \url{https://www.recordedfuture.com/vulnerability-disclosure-delay/}.

\bibitem[Lawless, 1987]{Lawless87}
Lawless, J.~F. (1987).
\newblock {N}egative {B}inomial and {M}ixed {P}oisson {R}egression.
\newblock {\em The Canadian Journal of Statistics}, 15(3):209--225.

\bibitem[Lesnoff and Lancelot, 2012]{aod}
Lesnoff, M. and Lancelot, R. (2012).
\newblock {aod}: {A}nalysis of {O}verdispersed {D}ata.
\newblock {R} {P}ackage {V}ersion 1.3. {A}vailable online in September 2017:
  \url{https://cran.r-project.org/web/packages/aod/index.html}.

\bibitem[Li and Sillanp\"a\"a, 2012]{LiSillanpaa12}
Li, Z. and Sillanp\"a\"a, M.~J. (2012).
\newblock {O}verview of {LASSO}-{R}elated {P}enalized {R}egression {M}ethods
  for {Q}uantitative {T}rait {M}apping and {G}enomic {S}election.
\newblock {\em Theoretical and Applied Genetics}, 125(3):419--435.

\bibitem[Massacci, 2017]{Massacci17}
Massacci, F. (2017).
\newblock {H}ow {D}o {Y}ou {K}now {T}hat {I}t {W}orks? {T}he {C}urses of
  {E}mpirical {S}ecurity {A}nalysis.
\newblock In Moore, T.~W., Probst, C.~W., Rannenberg, K., and van Eeten, M.,
  editors, {\em {A}ssessing {ICT} {S}ecurity {R}isks in {S}ocio-{T}echnical
  {S}ystems ({D}agstuhl {S}eminar 16461)}, volume~6, pages 77--78. Schloss
  Dagstuhl--Leibniz-Zentrum fuer Informatik, Dagstuhl.
\newblock Available online in September 2017:
  \url{http://drops.dagstuhl.de/opus/volltexte/2017/7039}.

\bibitem[McCullagh, 1983]{McCullagh83}
McCullagh, P. (1983).
\newblock {Q}uasi-{L}ikelihood {F}unctions.
\newblock {\em The Annals of Statistics}, 11(1):59--67.

\bibitem[Mell et~al., 2006]{Mell06}
Mell, P., Scarfone, K., and Romanosky, S. (2006).
\newblock {C}ommon {V}ulnerability {S}coring {S}ystem.
\newblock {\em IEEE Security \& Privacy}, 4(6):\text{85--89}.

\bibitem[Morrison et~al., 2017]{Morrison17}
Morrison, P.~J., Pandita, R., Xiao, X., Chillarege, R., and Williams, L.
  (2017).
\newblock {A}re {V}ulnerabilities {D}iscovered and {R}esolved {L}ike {O}ther
  {D}efects?
\newblock {\em Empirical Software Engineering}, 1--39.
\newblock Published online in September 2017.

\bibitem[Mu{\~n}oz-Gonz{\'a}lez et~al., 2017]{MunozGonzalez17}
Mu{\~n}oz-Gonz{\'a}lez, L., Sgandurra, D., Barr{\`e}re, M., and Lupu, E.~C.
  (2017).
\newblock {E}xact {I}nference {T}echniques for the {A}nalysis of {B}ayesian
  {A}ttack {G}raphs.
\newblock {\em IEEE Transactions on Dependable and Secure Computing}.
\newblock Published online in March 2017.

\bibitem[{NIST}, 2017a]{NVD17a}
{NIST} (2017a).
\newblock {NVD} {D}ata {F}eed and {P}roduct {I}ntegration.
\newblock {N}ational Institute of Standards and Technology~(NIST), Annually
  Archived CVE Vulnerability Feeds: Security Related Software Flaws, NVD/CVE
  XML Feed with CVSS and CPE Mappings (Version 2.0). Retrieved in 23 September
  2017 from: \url{https://nvd.nist.gov/download.cfm}.

\bibitem[{NIST}, 2017b]{NVD17b}
{NIST} (2017b).
\newblock {NVD} {F}requently {A}sked {Q}uestions.
\newblock {N}ational Institute of Standards and Technology~(NIST), Available
  online in November 2017: \url{https://nvd.nist.gov/general/faq}.

\bibitem[{NIST}, 2017c]{NVD17c}
{NIST} (2017c).
\newblock {V}ulnerability {M}etrics.
\newblock {N}ational Institute of Standards and Technology~(NIST), Available
  online in November 2017: \url{https://nvd.nist.gov/vuln-metrics}.

\bibitem[Ross et~al., 2017]{Ross17}
Ross, D.~M., Wollaber, A.~B., and Trepagnier, P.~C. (2017).
\newblock {L}atent {F}eature {V}ulnerability {R}anking of {CVSS} {V}ectors.
\newblock In {\em Proceedings of the Summer Simulation Multi-Conference
  (SummerSim 2017)}, pages 19:1 -- 19:12, Washington. ACM.

\bibitem[Ruohonen, 2017]{Ruohonen17TIR}
Ruohonen, J. (2017).
\newblock {C}lassifying {W}eb {E}xploits with {T}opic {M}odeling.
\newblock In {\em Proceedings of the 28th International Workshop on Database
  and Expert Systems Applications (DEXA 2017)}, pages 93--97, Lyon. IEEE.

\bibitem[Ruohonen et~al., 2016]{Ruohonen16GIQ}
Ruohonen, J., Hyrynsalmi, S., and Lepp\"anen, V. (2016).
\newblock {A}n {O}utlook on the {I}nstitutional {E}volution of the {E}uropean
  {U}nion {C}yber {S}ecurity {A}pparatus.
\newblock {\em Government Information Quarterly}, 33(4):746--756.

\bibitem[Ruohonen et~al., 2017a]{Ruohonen17COMSIS}
Ruohonen, J., Hyrynsalmi, S., and Lepp\"anen, V. (2017a).
\newblock {M}odeling the {D}elivery of {S}ecurity {A}dvisories and {CVEs}.
\newblock {\em Computer Science and Information Systems}, 14(2):537--555.

\bibitem[Ruohonen et~al., 2017b]{Ruohonen17IWSMMENSURA}
Ruohonen, J., Rauti, S., Hyrynsalmi, S., and Lepp\"anen, V. (2017b).
\newblock {M}ining {S}ocial {N}etworks of {O}pen {S}ource {CVE} {C}oordination.
\newblock In {\em Proceedings of the 27th International Workshop on Software
  Measurement and 12th International Conference on Software Process and Product
  Measurement (IWSM Mensura 2017)}, pages 176--188, Gothenburg. ACM.

\bibitem[Rydberg and Carkin, 2017]{Rydberg17}
Rydberg, J. and Carkin, D.~M. (2017).
\newblock {U}tilizing {A}lternate {M}odels for {A}nalyzing {C}ount {O}utcomes.
\newblock {\em Crime \& Delinquency}, 61(1):61--76.

\bibitem[Scarfone and Mell, 2009]{Scarfone09}
Scarfone, K. and Mell, P. (2009).
\newblock {A}n {A}nalysis of {CVSS} {V}ersion~2 {V}ulnerability {S}coring.
\newblock In {\em Proceedings of the 3rd International Symposium on Empirical
  Software Engineering and Measurement (ESEM 2009)}, pages 516--525, Lake Buena
  Vista. IEEE.

\bibitem[Shin et~al., 2015]{Shin15}
Shin, D.-H., Kim, H., and Hwang, J. (2015).
\newblock {S}tandardization {R}evisited: {A}~{C}ritical {L}iterature {R}eview
  on {S}tandards and {I}nnovation.
\newblock {\em Computer Standards \& Interfaces}, 38:152--157.

\bibitem[Stine et~al., 2017]{Stine17}
Stine, I., Rice, M., Dunlap, S., and Pecarina, J. (2017).
\newblock {A} {C}yber {R}isk {S}coring {S}ystem for {M}edical {D}evices.
\newblock {\em International Journal of Critical Infrastructure Protection}.
\newblock Published online in April 2017.

\bibitem[Vidaurre et~al., 2013]{Vidaurre13}
Vidaurre, D., Bielza, C., and Larra{\~n}aga, P. (2013).
\newblock {A} {S}urvey of {$L_1$} {R}egression.
\newblock {\em International Statistical Review}, 81(3):361--387.

\bibitem[Wang et~al., 2012]{WangGuo12}
Wang, J.~A., Guo, M., Wang, H., and Zhou, L. (2012).
\newblock {M}easuring and {R}anking {A}ttacks {B}ased on {V}ulnerability
  {A}nalysis.
\newblock {\em Information Systems and e-Business Management}, 10(4):455--490.

\bibitem[Wei and Lovegrove, 2013]{WeiLovegrove13}
Wei, F. and Lovegrove, G. (2013).
\newblock {A}n {E}mpirical {T}ool to {E}valuate the {S}afety of {C}yclists:
  {C}ommunity {B}ased, {M}acro-{L}evel {C}ollision {P}rediction {M}odels
  {U}sing {N}egative {B}inomial {R}egression.
\newblock {\em Accident Analysis \& Prevention}, 61:129--137.

\bibitem[White, 1980]{White80}
White, H. (1980).
\newblock {A} {H}eteroskedasticity-{C}onsistent {C}ovariance {M}atrix
  {E}stimator and a {D}irect {T}est for {H}eteroskedasticity.
\newblock {\em Econometrica}, 80(4):\text{817--838}.

\bibitem[Younis and Malaiya, 2015]{YonisMalaiya15}
Younis, A.~D. and Malaiya, Y.~K. (2015).
\newblock {C}omparing and {E}valuating {CVSS} {B}ase {M}etrics and {M}icrosoft
  {R}ating {S}ystem.
\newblock In {\em Proceedings of the IEEE International Conference on Software
  Quality, Reliability and Security (QRS 2015)}, pages 252--261, Vancouver.
  IEEE.

\bibitem[Zeileis, 2004]{Zeileis04}
Zeileis, A. (2004).
\newblock {E}conometric {C}omputing with {HC} and {HAC} {C}ovariance {M}atrix
  {E}stimators.
\newblock {\em Journal of Statistical Software}, 11(10):\text{1--17}.

\bibitem[Zhu et~al., 2017]{ZhuCao17}
Zhu, X., Cao, C., and Zhang, J. (2017).
\newblock {V}ulnerability {S}everity {P}rediction and {R}isk {M}etric
  {M}odeling for {S}oftware.
\newblock {\em Applied Intelligence}, 47(3):828--836.

\end{thebibliography}

\end{document}